# On 'Bell violation using entangled photons without the fair-sampling assumption'

Johannes Kofler[1], Sven Ramelow[2,3], Marissa Giustina[2,3], Anton Zeilinger[2,3]

[1] Max Planck Institute of Quantum Optics (MPQ), Hans-Kopfermannstraße 1, 85748 Garching, Germany.
[2] Institute for Quantum Optics and Quantum Information – Vienna (IQOQI), Austrian Academy of Sciences, Boltzmanngasse 3, 1090 Vienna, Austria.
[3] Vienna Center for Quantum Science and Technology (VCQ), Faculty of Physics, University of Vienna, Boltzmanngasse 5, 1090 Vienna, Austria.

A recent experiment [1] presented, for photons, the first violation of a Bell inequality closing the fair-sampling loophole, i.e., without having to assume that the sample of measured photons fairly represents the entire ensemble. In this note, we discuss a detailed quantum mechanical model for the experimental data. Within experimental error the experiment agrees with quantum mechanical prediction. We also discuss the effects of drifting laser intensity and show that it could not have mimicked a violation of the Bell inequality. Therefore, the experiment was immune to the "production-rate loophole."

## 1. A quantum mechanical model for the data of Ref. [1]

Tests of Bell's inequalities and their variants are primarily tests of the basic assumptions underlying the derivation of the inequalities. These are generally seen as realism and locality ("local realism"), though other assumptions also enter, some very subtly. A violation of the inequality may then be interpreted as ruling out a local realistic worldview, possibly with the caveat that some loopholes might still be open due to such additional assumptions. Whether the data can be explained by quantum physics is a question independent of the validity of local realism. While the success of quantum mechanics certainly suggests that such a quantitative explanation must be possible, it is irrelevant for the question of local realism. However, certainly a quantum mechanical explanation of the data adds to the credibility of an experiment.

Any experimental data that violates a CH inequality [2] like Eberhard's inequality [3] rules out a local realistic explanation exploiting the detection loophole. Although, as mentioned above, it is logically unnecessary to compare the data in Ref. [1] to the quantum mechanical predictions, the experiment was modeled quantum-mechanically nevertheless. Excellent agreement with the observed data was found and reported (albeit without details) in Ref. [1]. In light of recent remarks (e.g. [4] as well as direct questions from some colleagues), we believe there is some interest in the further details of the theoretical model and how it actually predicts the recorded data of Ref. [1].

In the following, we will use the notation and terminology of Ref. [1]. The ideal Eberhard states take the form $|\psi_r\rangle = (1+r^2)^{-1/2}(|HV\rangle + r|VH\rangle)$, with $0 < r < 1$ and H (V) denoting horizontal (vertical) polarization of Alice's and Bob's photons.

In a first step, one needs to account for the impurity of the state that is actually produced in the source. This mixed state can be approximated very well by using the following parameterization:

$$\hat{\rho} = \frac{1}{1+r^2} \begin{pmatrix} 0 & 0 & 0 & 0 \\ 0 & 1 & Vr & 0 \\ 0 & Vr & r^2 & 0 \\ 0 & 0 & 0 & 0 \end{pmatrix} \qquad (1)$$

in the standard H/V basis. The off-diagonal terms are damped by a positive real factor $V < 1$. In reality that factor will be complex and there will be no elements of zero value in any realistic density matrix. These points will be addressed later. They are negligible to the present analysis.

For $r = 1$, $V$ can be understood as the visibility in the diagonal basis. The visibility in the H/V basis remains perfect in this model even for $V < 1$. This is a consequence of the experimental scheme with which polarization-entanglement is created in the source, namely as a superposition of pairs each created either in the product state $|HV\rangle$ or in $|VH\rangle$. From the calibration measurements that are used to set the state the parameter $V$ can be estimated to around 96.5 %. (Note that the reported 97.5 % in Ref. [1] is a misprint and should read 96.5 %.)

The total rate $R_0$ of produced pairs was determined by a measurement of all combinations in the H/V basis ($\alpha = 0°, 90°$; $\beta = 0°, 90°$) yielding around 80 700 produced pairs per second. From the H/V measurement, it is possible to directly determine $r$ to around 0.297 as well as the total arm efficiencies for Alice ($\eta_A \approx 73.77\ \% \pm 0.07\ \%$) and Bob ($\eta_B \approx 78.59\ \% \pm 0.08\ \%$). The expected ordinary "$o$" singles and coincidence counts (not yet corrected for accidental coincidences and background and thus labeled by a tilde sign) within measurement time $T$ for measurement angles $\alpha_i$ and $\beta_j$ are given by:

$$\widetilde{S}_o^A(\alpha_i) = \eta_A R_0 T \operatorname{Tr}[\hat{\rho}(\hat{P}_A(\alpha_i) \otimes \hat{I}_B)] \tag{2}$$

$$\widetilde{S}_o^B(\beta_j) = \eta_B R_0 T \operatorname{Tr}[\hat{\rho}(\hat{I}_A \otimes \hat{P}_B(\beta_j))] \tag{3}$$

$$\widetilde{C}_{oo}(\alpha_i, \beta_j) = \eta_A \eta_B R_0 T \operatorname{Tr}[\hat{\rho}(\hat{P}_A(\alpha_i) \otimes \hat{P}_B(\beta_j))] \tag{4}$$

Here, $\hat{P}$ are projectors on the corresponding polarization states of Alice and Bob and $\hat{I}$ denotes the identity operator.

In a second step, the singles counts (2) and (3) need to be corrected for the background (dark) counts, which are detected at a rate $\zeta$ of about 10 events per second. This leads to an estimate for the experimentally observed singles counts during time $T$:

$$S_o^A(\alpha_i) = \widetilde{S}_o^A(\alpha_i) + \zeta T \tag{5}$$

$$S_o^B(\beta_j) = \widetilde{S}_o^B(\beta_j) + \zeta T \tag{6}$$

Moreover – and especially relevant for the $\alpha_2\beta_2$ setting combination – the coincidence count expression (4) needs to be corrected to reflect the presence of accidental coincidences $C_{oo}^{\mathrm{acc}}(\alpha_i, \beta_j)$. An accidental coincidence occurs when photons not belonging to the same pair are detected within the coincidence window $\tau_C = 180$ ns.

The accidental coincidence counts per measurement time $T$ can be accurately estimated for each setting combination given the corresponding singles counts, the coincidence-to-singles ratios and the coincidence window:

$$C_{oo}^{\mathrm{acc}}(\alpha_i, \beta_j) = S_o^A(\alpha_i) S_o^B(\beta_j) \frac{\tau_C}{T} \left(1 - \frac{\widetilde{C}_{oo}(\alpha_i, \beta_j)}{S_o^A(\alpha_i)}\right)\left(1 - \frac{\widetilde{C}_{oo}(\alpha_i, \beta_j)}{S_o^B(\beta_j)}\right) \tag{7}$$

The last two factors (that are in addition to the usual formula used for low arm efficiencies) account for the fact that an accidental coincidence can only occur in cases where the *real* coincidence is not detected. This is especially relevant for the first three setting combinations ($\alpha_1\beta_1$, $\alpha_1\beta_2$, and $\alpha_2\beta_1$). Note that a detailed analysis of accidental coincidences can be found in the appendix of Ref. [5].

We can now write the estimate for the experimentally observed coincidence counts:

$$C_{oo}(\alpha_i, \beta_j) = \widetilde{C}_{oo}(\alpha_i, \beta_j) + C_{oo}^{\mathrm{acc}}(\alpha_i, \beta_j) \tag{8}$$

Using the above expressions (1) to (8) and input parameters $r = 0.297$, $V = 96.5\ \%$, $\eta_A = 73.77\ \%$, $\eta_B = 78.59\ \%$, $R_0 = 80\ 700$ Hz, $T = 300$ s, ($N = R_0 T = 24.21 \cdot 10^6$,) $\zeta = 10$ Hz, $\tau_C = 180$ ns, $\alpha_1 = 85.6°$, $\alpha_2 = 118.0°$, $\beta_1 = -5.4°$, and $\beta_2 = 25.9°$, the experimentally observed coincidence and singles counts are very well reproduced with deviations which are partially even within a purely statistical (1-sigma Poissonian) error (see table 1). This is the very good agreement that was reported in Ref. [1].

|  | $C_{oo}(\alpha_1,\beta_1)$ | $C_{oo}(\alpha_1,\beta_2)$ | $C_{oo}(\alpha_2,\beta_1)$ | $C_{oo}(\alpha_2,\beta_2)$ | $S_o^A(\alpha_1)$ | $S_o^B(\beta_1)$ | $J$ |
|---|---|---|---|---|---|---|---|
| Exp. Data | 1 069 306 | 1 152 595 | 1 191 146 | 69 749 | 1 522 865 | 1 693 718 | –126 715 |
| Qm. Model | 1 068 886 | 1 152 743 | 1 192 489 | 68 694 | 1 538 766 | 1 686 467 | –120 191 |
| Deviation | –0,04 % | 0,01 % | 0,11 % | –1,51 % | 1,04 % | –0,43 % | 5,15 % |

Table 1. Comparison of the experimental data from Ref. [1] with the quantum mechanical model described in the main text for the coincidences and singles counts. The deviation for the individual values in the table is defined as the relative difference of the model from the data.

The accidental coincidences in the $\alpha_1,\beta_1$ setting combination contribute about 0.02 % of the observed coincidences, for $\alpha_1,\beta_2$ and $\alpha_2,\beta_1$ around 0.1 %, and for the $\alpha_2,\beta_2$ setting combination approximately 18 %. The contribution in the $\alpha_2,\beta_2$ setting combination contributes positively to the Eberhard value and far outweighs (both relatively and absolutely) the contribution in all other settings combined. This implies that accidental coincidences in fact made it harder to violate the bound of the Eberhard inequality in the experiment [1].

Finally, we believe the remaining small deviations of the observed rates from the ones predicted by the above model are caused by residual imperfections, most notably:
- The visibility in the H/V basis is not perfect but "only" around 99.8 % (because of small imperfections in the alignment and the polarization optics in the source, especially the polarizing beam splitter and dual wave-plate). As a consequence, the diagonal of the actual quantum state cannot contain perfectly vanishing elements.
- Similar small deviations from the state (1) are expected in general: all off-diagonal elements are never perfectly zero, and also $r$ is expected to have a non-vanishing complex phase factor.
- Calibration errors on the order of a few tenths of a degree of the wave-plates used to set the measurement settings can lead to changes in the number of counts.
- Because of imperfections in alignment, the total arm efficiencies can be slightly different for the two pump directions of the Sagnac source. This can lead to a small, local setting dependence of the total arm efficiencies $\eta_A$ and $\eta_B$.
- Intensity drifts of the laser cause temporal variations of the pair production rate (see chapter 2).

Note that all these effects do not in any way influence the validity of the local realism violation, but only affect the accuracy with which the observed data can be modeled quantum mechanically as described above.

## 2. The "production-rate loophole"

In an experimental test of Eberhard's or any other CH-type inequality, it is important to check whether the number of emitted particle pairs depends on the chosen measurement settings. If, for instance, the intensity of the pump laser and thus of the emitted particles strongly drops for those time intervals when measurements are made with settings $\alpha_2,\beta_2$, then a violation of the inequality can be simulated within local realism. We will call this type of loophole the "production-rate loophole".

In contrast to the suggestion in Refs. [6,7], this loophole *cannot* be exploited in the experiment [1] to explain the observed results local-realistically. This can be seen by the following quantitative analysis. Let us start by considering the complete set of experimental data for the accumulated singles counts under all setting combinations (table 2).

The relative deviations $\Delta$ of singles counts for the same setting (between –0.12 % and –0.36 %) are about two to five times larger than expected from purely statistical fluctuations. The latter are quantified by the inverse square root of the counts, and are approximately 0.08 % for the smaller count numbers $S_o^A(\alpha_1)$ and $S_o^B(\beta_1)$, and about 0.05 % for the larger count numbers $S_o^A(\alpha_2)$ and $S_o^B(\beta_2)$. The size of the deviations indicates that there were indeed some additional effects like residual drifts in the laser intensity that did not completely average out during the measurement rounds. Moreover, the deviations were indeed such that they helped in violating the Eberhard inequality, because the production rate was smallest for setting $\alpha_2,\beta_2$. However, the effects are quantitatively far too small to account for a local realistic explanation. We will show this in more detail by analyzing both the total counts and the individual measurement rounds.

| $\alpha,\beta$ | $S_o^A$ | $\Delta$ | $S_o^B$ | $\Delta$ |
|---|---|---|---|---|
| $\alpha_1,\beta_1$ | 1 526 617 | | 1 699 881 | |
| $\alpha_1,\beta_2$ | 1 522 865 | –0,25 % | 4 515 782 | |
| $\alpha_2,\beta_1$ | 4 735 046 | | 1 693 718 | –0,36 % |
| $\alpha_2,\beta_2$ | 4 729 369 | –0,12 % | 4 507 497 | –0,18 % |

Table 2. Alice's and Bob's accumulated singles counts in 300 seconds for all four setting combinations. Alice's $\alpha_1$ singles counts relatively differ by –0,25 % in the two setting combinations $\alpha_1,\beta_1$ and $\alpha_1,\beta_2$ (black). Similarly, her $\alpha_2$ singles counts differ by –0,12 % (green). Bob's relative differences are –0,36 % in his $\beta_1$ counts (red) and –0,18 % in his $\beta_2$ counts (blue).

In Ref. [1], one minute of data was recorded in each setting as the settings were cycled for multiple rounds. Each measurement round consisted of switching through the combinations $\alpha_1,\beta_1$, $\alpha_1,\beta_2$, $\alpha_2,\beta_2$, $\alpha_2,\beta_1$, such that every combination change required the alteration of only one setting. The whole experiment consisted of 5 such rounds. In its first 6 rows, table 3 shows the complete set of data for the singles and coincidence counts as well as the resulting Eberhard values $J$ for all 5 measurement rounds. The accumulated data is shown at the bottom of the table.

Column 7 shows a correction factor denoted as $f$, which is proportional to the average intensity (average production rate) during each different setting combination. (Again, the deviations between the setting combinations are a few times larger than expected from purely statistical fluctuations.) Since intensity drifts within a chosen setting combination are irrelevant, the values of $f$ can be used to normalize any drift out of the data within a given round. The corrected data (not shown) is what we would expect to have observed if the production rate had been constant. Using the corrected data we can calculate a corrected $J$-value $J'$, which is the $J$-value expected for a constant production rate.

For every round, all correction factors are defined relatively between two measured values. We compute $f$ in the sequence of the setting combinations, denoting $f$ for $\alpha_i,\beta_j$ by $f_{ij}$. Then $f_{12}$ can be expressed as a multiple of $f_{11}$ by comparing Alice's $\alpha_1$ singles counts, $f_{22}$ is expressed as a multiple of $f_{12}$ by comparing Bob's $\beta_2$ singles counts, and $f_{21}$ is expressed in terms of $f_{22}$ by comparing Alice's $\alpha_2$ singles counts. We note that this order is not the only option. We could, e.g., also compute $f_{21}$ from $f_{11}$ by comparing Bob's $\beta_1$ singles counts, as can easily be seen in table 2. Different "paths" lead to slightly different results because of the intrinsic statistical (Poissonian) fluctuations between Alice's and Bob's counts for one and the same combination, which already indicates that the choice of path cannot be of statistical significance.

Having computed the relationships between correction factors, it is necessary to set a "baseline" production rate. Due to their set relationships, this fixes the values of all four $f_{ij}$. To be conservative, we always choose the best case for the local realist who tries to maximize $J$. This best case choice is to set the smallest $f_{ij}$ as the baseline (100 %), such that the other three $f_{ij}$ are > 100 %, allowing him the maximal possible suppression of the negative $J$-value. Then each of the singles and coincidence counts is *divided* by its corresponding $f$ factor in order to get normalized counts (not shown), which then result in the 5 adapted values of the Eberhard inequality, denoted by $J'$ in table 3. Note that we perform normalizations only within the rounds, not between them. In rounds 2, 3, and 5 the drifts were detrimental to the violation of the inequality. Only in rounds 1 and 4 does the normalization lead to an increase from $J$ to $J'$ (that is, $J < J'$). Nevertheless, the accumulated $J'$ is larger than the accumulated $J$.

The original $J$-values have mean –25 343, standard deviation 1 503, and accumulated value –126 715. The adapted $J'$-values have mean –24 626, standard deviation 1 243, and accumulated value –123 132. The statistical significance (in terms of accumulated or mean Eberhard value divided by standard deviation) of the violation is actually increased by the normalization procedure. This is because the normalization procedure takes away intensity drifts within the rounds. (One might argue that the local realist will not perform normalizations in rounds 2, 3, and 5 where it actually would weaken his point. Even allowing him this procedure, he would obtain the following values: mean –24 215, standard deviation 893, and accumulated value –121 076.)

The normalization technique is also applied directly to the total counts (bottom of table 3), leading to an adapted $J$ of –123 412. The deviation from the value –123 132 above is within statistical fluctuation and stems mainly from the fact that we allowed for normalizations with respect to different combinations in the different rounds. (If, e.g., we *always* normalize with respect to $\alpha_1,\beta_1$, we get the following results for the 5 measurement rounds: mean –24 787, standard deviation 1 098, and accumulated value –123 935; as well as –123 943 when applied directly to the total counts.) We remark again that different "paths" in the normalizations slightly change all results.

| Round | $\alpha,\beta$ | $S_o^A$ | $S_o^B$ | $C_{oo}$ | $J$ | $f$ | $J'$ |
|---|---|---|---|---|---|---|---|
| 1 | $\alpha_1,\beta_1$ | 308 131 | 341 484 | 215 282 | | 102,07% | |
|   | $\alpha_1,\beta_2$ | 302 394 | 897 934 | 228 605 | | 100,17% | |
|   | $\alpha_2,\beta_2$ | 940 904 | 896 442 | 14 501 | | 100,00% | |
|   | $\alpha_2,\beta_1$ | 945 152 | 337 158 | 238 151 | –27 985 | 100,45% | –24 193 |
| 2 | $\alpha_1,\beta_1$ | 303 988 | 338 929 | 212 953 | | 100,00% | |
|   | $\alpha_1,\beta_2$ | 304 593 | 900 646 | 231 168 | | 100,20% | |
|   | $\alpha_2,\beta_2$ | 943 776 | 900 898 | 13 861 | | 100,23% | |
|   | $\alpha_2,\beta_1$ | 946 507 | 339 996 | 239 361 | –25 032 | 100,52% | –25 727 |
| 3 | $\alpha_1,\beta_1$ | 305 770 | 341 446 | 214 545 | | 100,00% | |
|   | $\alpha_1,\beta_2$ | 306 556 | 909 078 | 231 487 | | 100,26% | |
|   | $\alpha_2,\beta_2$ | 954 277 | 907 159 | 13 538 | | 100,05% | |
|   | $\alpha_2,\beta_1$ | 956 094 | 341 548 | 239 889 | –24 279 | 100,24% | –24 717 |
| 4 | $\alpha_1,\beta_1$ | 307 790 | 342 499 | 214 853 | | 101,33% | |
|   | $\alpha_1,\beta_2$ | 307 340 | 910 570 | 231 871 | | 101,18% | |
|   | $\alpha_2,\beta_2$ | 949 608 | 905 976 | 14 237 | | 100,67% | |
|   | $\alpha_2,\beta_1$ | 943 247 | 336 659 | 236 109 | –24 597 | 100,00% | –22 750 |
| 5 | $\alpha_1,\beta_1$ | 300 938 | 335 523 | 211 673 | | 100,00% | |
|   | $\alpha_1,\beta_2$ | 301 982 | 897 554 | 229 464 | | 100,35% | |
|   | $\alpha_2,\beta_2$ | 940 804 | 897 022 | 13 612 | | 100,29% | |
|   | $\alpha_2,\beta_1$ | 944 046 | 338 357 | 237 636 | –24 822 | 100,63% | –25 745 |
|   | Sum | | | | –126 715 | | –123 132 |
|   | Mean | | | | –25 343 | | –24 626 |
|   | St. dev. | | | | 1 503 | | 1 243 |
| Total | $\alpha_1,\beta_1$ | 1 526 617 | 1 699 881 | 1 069 306 | | 100,43% | |
|   | $\alpha_1,\beta_2$ | 1 522 865 | 4 515 782 | 1 152 595 | | 100,18% | |
|   | $\alpha_2,\beta_2$ | 4 729 369 | 4 507 497 | 69 749 | | 100,00% | |
|   | $\alpha_2,\beta_1$ | 4 735 046 | 1 693 718 | 1 191 146 | –126 715 | 100,12% | –123 412 |

Table 3. The complete set of data for the singles and coincidence counts as well as the resulting Eberhard values for all 5 measurement rounds. Note the different sequence of setting combinations compared to table 2, now matching the experimental procedure. See text for further details.

The above analysis shows that intensity deviations were not only present in the experiment, they exceeded purely statistical fluctuations and even decreased the measured Eberhard value. However, the effects were by far too small to allow a local realistic explanation. Rather, although $J'$ becomes larger than $J$ and thus closer to the local realistic bound, the statistical significance (in terms of Eberhard value divided by standard deviation) of the violation is even increased, when normalization is taken into account. We conclude that no loophole with respect to drifting laser intensity was left open in the experiment [1].

When closing the production-rate loophole, using randomness and regular switching for the setting choices [6,7] might be of practical use as it helps to equalize the normalization factors within the statistical fluctuations. However, randomness and switching are irrelevant from a logical point of view. To close the loophole – which by definition only refers to the possibility of the production rate being (conspiratorially) correlated with the settings – it is sufficient to normalize the data, even in cases of significantly different intensities for the different setting combinations. The situation is very different for closing the freedom-of-choice loophole [8,9], where randomness and switching are indeed of utmost importance. However, it is essential to remark that to close the latter loophole the settings must be randomly chosen not only at regular instances but for every particle pair. (Unlike [9], neither the experiment in [1] nor in [6,7] was able to ensure the freedom-of-choice condition.) This guarantees the independence of every setting choice from the corresponding particle pair's hypothetical hidden variables and thus automatically ensures independence from potentially changing production rates.

**Conclusion**

In this short note, we have presented a quantum mechanical model that accurately describes the experimental data reported in Ref. [1]. Moreover, we have demonstrated that the experiment was not vulnerable to the production-rate loophole. A detailed analysis regarding the coincidence-time loophole [10] will be presented elsewhere [11].

**Acknowledgments:** We acknowledge comments from and discussions with Daniel Greenberger, Michael A. Horne, Emanuel Knill, Sae Woo Nam, Rupert Ursin, and Bernhard Wittmann. This work was supported by the Austrian Science Fund (FWF) under projects SFB F4008 and CoQuS, by the European Commission under grants Q-ESSENCE (no. 248095), SIQS (No. 600645), and the John Templeton Foundation.